\documentclass{elsart}

\begin{document}

\begin{frontmatter}

\title{Stochastic group selection model for the evolution of altruism}

\author{ Ana T. C. Silva}
and
\author{J. F. Fontanari \thanksref{email}} 
\address{Instituto de F\'{\i}sica de S\~ao Carlos,
Universidade de S\~ao Paulo,
Caixa Postal 369,
13560-970 S\~ao Carlos, SP, Brazil}

\thanks[email]{Corresponding author. E-mail:
 fontanari@ifsc.sc.usp.br}

\begin{abstract}

We study numerically and analytically a stochastic group 
selection model in which 
a population of asexually reproducing individuals, each of which can 
be either altruist or non-altruist, is subdivided into $M$ 
reproductively isolated groups (demes) of  size $N$. The cost associated
with being altruistic is modelled by assigning the fitness
$1- \tau$, with $\tau \in [0,1]$, to the altruists and the fitness $1$
to the non-altruists. 
In the case that the altruistic 
disadvantage $\tau$ is not too large, we show that the finite $M$
fluctuations are small
and practically do not alter the deterministic results
obtained for $M \rightarrow \infty$. However, for large $\tau$
these fluctuations greatly increase the instability of the
altruistic demes to mutations. These results may be relevant
to the dynamics of parasite-host systems and, in particular,
to explain the importance of mutation in the evolution of
parasite virulence.

\end{abstract}


\end{frontmatter}

%
\section{Introduction}
%

Despite the scarcity of empirical facts  supporting group selection
as a relevant evolutive force in nature, the mathematical problems 
involved in its  modeling  have kept a recurrent theoretical
interest on this controversial theory \cite{Levins,book}.  
Group selection is based on an analogy
between individuals (or genes) and reproductively isolated 
subpopulations, termed demes. If the extinction of demes
occurs at a rate depending on their composition, then such extinctions 
will favor the existence of individuals that increase the probability 
of survival of the deme they belong to. 
In the case that these individuals are disfavored by the usual 
selection at the individual level,
group selection will oppose individual selection, and so
it has been advanced as an explanation for 
the existence of altruistic traits in nature.
Such a trait is defined as one that is detrimental to
the fitness of the individual who expresses it, but that confers an
advantage on the group of which that individual is a member.

The standard mathematical framework to study group selection was 
proposed by Levins more than 20 years ago \cite{Levins,book}.
The key ingredients are the differential  survival probability 
favoring demes with a large number of altruists, and the subsequent 
recolonization of the extinguished  demes by the surviving ones.
In fact, this is practically the only generally accepted mechanism 
to produce group selection in nature (see \cite{Roberta,DPS} for
an alternative proposal). However,
the mathematical complexity of Levins' formulation, based on 
a nonlinear integral partial differential equation, as well as 
the need for too restrictive assumptions,  have motivated
the proposal and study of a variety of discrete time
versions of  Levins' model \cite{Kilmer,Eshel,Aoki,Ana}. 
These analyses have concentrated mainly on the deterministic 
regime, in which the number of demes $M$ is infinite, though the deme
size $N$ (i.e., the number of individuals in each deme) is finite.
In the absence of mutation,
the finitude of $N$ is crucial to guarantee the fixation
through random drift of the altruistic trait within some demes.
The undesirable feature of considering $M$ finite  as 
well, besides obliterating the possibility of an analytical solution 
to the problem, is that the fluctuations occurring during the extinction 
process will ultimately lead to the complete extinction of the 
population. This is a consequence of the sequential procedure that 
considers first the extinction of  demes and then the recolonization of 
the extinct demes by the surviving ones.
In this paper we modify the sequential extinction-recolonization 
procedure so as to avoid global extinction, allowing thus the
numerical and analytical study of the effects of a finite population
on the steady-state of the metapopulation (i.e., the population of demes). 
More pointedly, once a deme is extinct we immediately assign one of the
$M-1$ surviving demes to replace it, although the effective replacement 
of the extinct demes will take place only after all demes have passed
the extinction stage. This replacement or recolonization occurs
simultaneously for all demes.

Our goal is to study the effects of the fluctuations due to the finitude of
the population on the stability of the altruistic state predicted
by Eshel in the deterministic regime \cite{Eshel}. 
The remainder of this paper is organized as follows. 
In section 2 we describe the events that comprise
the life cycles of the individuals in the metapopulation
 and present the stochastic dynamics
governing the time evolution of the metapopulation. A mean-field recursion
equation is derived and its validity discussed in section 3.
The results of the simulations as well as those of the mean-field approximation
are presented and analyzed in section 4.  
Finally, in section 5 we present some concluding remarks and, in
particular, point out the relevance of our results to the dynamics
of parasite-host systems.

%
\section{Model}
%

The metapopulation is composed of $M$ demes, each of which being 
composed of $N$ haploid, asexually reproducing  individuals. 
An individual can be either altruist or non-altruist. The
cost associated with being altruistic is modelled by assigning the 
reproductive rate
$1 -\tau$, with $\tau \in [0,1]$, to the altruists and 
the reproductive rate $1$ to the non-altruists. 
The demes are classified according to the number of altruistic
individuals they have, so that there are $N+1$ different types of 
demes, labeled by the integers $i = 0,1, \ldots, N$. In each generation
the metapopulation is described by the vector
${\bf n} = \left ( n_0, n_1, \ldots, n_N \right ) $, where $n_i$ is
the number of demes of type $i$, so that $\sum_i n_i = M$.
The life cycle (i.e., one generation) consists of the following
events, which will be discussed in detail in the sequel: extinction,
recolonization, reproduction, and mutation. 

\subsection{Extinction and Recolonization}

Within the differential extinction framework, we define the probability
that a deme of type $i$ survives  extinction, $\alpha_i$,  by 
\begin{equation}\label{alpha_i}
\alpha_i   =  \left \{ \begin{array}{ll}
  \half \left ( 1 + i/i_c \right ) & \mbox{~~~~~if $i < i_c$ } \\
                  1 & \mbox{~~~~~otherwise} ,
      \end{array}
      \right.  
\end{equation}
where $i_c = 0, 1, \ldots, N$ is a parameter measuring the intensity
of the group selection pressure. The larger the number of altruists
in a deme, the larger its chance of surviving extinction.
Once a deme is extinct, a randomly chosen
deme among the $M-1$ surviving ones will immediately be assigned
to replaced it. This contrasts with the standard
modelling in which recolonization takes places only after all 
demes have passed the extinction procedure.
Hence, given ${\bf n}$, the probability
that a deme of type $j$ changes to a deme of type $i$, denoted by
$E_{ij}$, is simply

\begin{equation}\label{sur}
E_{ij}   =  \left \{ \begin{array}{ll}
 \alpha_i +  \left( 1 - \alpha_i \right ) \left ( n_i - 1\right )/
\left ( M-1 \right ) 
    & \mbox{~~~~~if $i = j$ } \\
                  \left ( 1 - \alpha_j \right ) n_i/\left ( M-1 \right ) 
 & \mbox{~~~~~if $i \neq j$} .
      \end{array}
      \right.  
\end{equation}
As expected, $\sum_i E_{ij} = 1 ~\forall j$. We note that the transition 
matrix ${\bf E}$ depends on ${\bf n}$ and so it changes as the population 
evolves. The conjunction of the extinction and recolonization procedures 
is termed interdemic selection since the correlation between the 
elements belonging to a same column  of ${\bf E}$
yields an effective, indirect interaction between the demes.

\subsection{Reproduction}

The reproduction process occurs inside
the demes and hence is termed intrademic selection. Since the size
of the demes is fixed and finite ($N$), random drift occurs.
Following Wright's classical model \cite{Wright} we assume that the 
number of offspring that an individual contributes to the new generation 
is  proportional to its relative reproductive rate. Thus, 
the probability that a deme of type $j$  changes to a deme
of type $i$ is written as
\begin{equation}\label{R}
R_{ij} = \left ( \! \! \begin{array}{c} N \\ i 
\end{array}
\! \! \right ) \, w_j^{i} \,
 \left ( 1 - w_j \right)^{N - i} ,
\end{equation}
where
\begin{equation}
w_j = \frac{j \left( 1-\tau \right )}{N - j \tau}
\end{equation}
is  the relative reproductive rate of the subpopulation of altruists
in a deme of type $j$. We note that $\sum_i R_{ij} = 1 ~\forall j$
and $\sum_i i R_{ij} = N w_j$. In the absence of mutations, the 
random drift inherent to the reproduction process will prevent the
existence of mixed demes, i.e., a deme will be either of type $N$
(only altruists) or of type $0$ (only non-altruists). As a result,
a choice of the parameter $i_c$ different from $0$ or $N$ in the 
definition of the survival probability  $\alpha_i$ will have practically no
effect on the metapopulation evolution.

\subsection{Mutation}

To include mutation into the model, we must descend to the level
of the genes that determine the characteristics of the individuals.
In particular, we assume that two alleles, say  $A$ or $B$, at a single 
locus determine whether a given
individual is  altruist or non-altruist, respectively.
Since the replication of a gene may not be perfect, we introduce the 
mutation rate $u \in [0,1/2]$, which gives the
probability that the allele  $A$ mutates to  $B$ and vice-versa.
Hence the probability  that a deme of type $j$ changes to a
deme of type $i$ due to mutations of its members
is given by 
\begin{equation}\label{U}
U_{ij} = \sum_{l = l_l}^{l_u} \left ( \! \! \begin{array}{c} j \\ l 
\end{array}
\! \! \right ) \, \left ( \! \! \begin{array}{c} N - j \\ i - l
\end{array} \! \! \right ) \, u^{i + j - 2l} \,
 \left ( 1 - u \right)^{N - i - j + 2l} ,
\end{equation}
where $l_l = \mbox{max} \left ( 0,i+j-N \right)$ and
$l_u = \mbox{min} \left ( i,j  \right)$. Clearly, 
$\sum_i U_{ij} = 1~ \forall j$ and $\sum_i i U_{ij} =
N u + j \left ( 1 - 2u \right ) $. 

\subsection{Stochastic dynamics}

Given a population characterized by the vector ${\bf n}$
which will change into a new population characterized by
${\bf n}'$ due to a generic transition matrix ${\bf T}$ 
($\sum_i T_{ij} = 1 ~~\forall j $),  the stochastic dynamics is
defined by the conditional probability 
distribution $P_T \left ( {\bf n}' | {\bf n} \right )$.
To evaluate this quantity, it is more convenient to introduce the set
of integers $\{ b_{ij} \} $, where $b_{ij}$ stands for the 
number of demes of type $j$ that have changed to a deme of type $i$. 
Hence $n_j = \sum_i b_{ij}$  and $n_i' = \sum_j b_{ij}$, so that
given the set  $\{ b_{ij} \}$, the vector ${\bf n}'$ can be readily 
determined.
In fact, given $ n_j$ the conditional  probability  distribution
of ${\bf b}_j = \left ( b_{0j}, b_{1j},\ldots, 
b_{N j} \right )$ is  a multinomial
\begin{equation}\label{P_gen}
P_T \left ( {\bf b}_j | n_j \right ) = \frac{n_j !}{b_{0j}! \, b_{1j}!
\ldots b_{N j} !} \, T_{0j}^{b_{0j}} \, T_{1j}^{b_{1j}} \ldots
T_{N j}^{b_{N j}}
\end{equation}
for  $j=0,\ldots, N$. Clearly, the random variables $b_{kj}$ 
and $b_{li}$ are 
statistically independent for $i \neq j$, regardless of the values 
assumed by the indices $k$ and $l$.
We must emphasize that since the transition matrix ${\bf E}$, which governs 
the extinction and recolonization procedures, depends explicitly 
on ${\bf n}$, this
dynamics must be applied in parallel (simultaneously) to all demes.

The dynamics proceeds as follows. Given the population vector in
generation $t$, denoted by ${\bf n}^t$, first we consider the 
extinction-recolonization event and generate the conditional 
probability distributions 
$P_E \left ( {\bf b}_j |  n_j^t \right ) ~\forall j$.
The choice of $(N+1)^2$  uniformly distributed random numbers allows
the determination of the set $\{ b_{ij} \}$ and, consequently,
of the new population  vector ${\bf n}'$. Next, given ${\bf n}'$ we 
repeat this procedure for the reproduction event, then generating
the population vector ${\bf n}''$. Finally, the same procedure is repeated
again for the mutation  event, leading from ${\bf n}''$ to 
${\bf n}^{t+1}$, which thus completes the life cycle.


\section{Expectations}


The (conditional) expected value 
of the number of demes of type $i$ after a life cycle given the
population vector ${\bf n}^t$ in generation $t$ is defined by
\begin{equation}
\langle n^{t+1}_i \rangle = \sum_{{\bf k}, {\bf l},{\bf m}} 
k_i P_U \left ( {\bf k} | {\bf l} \right )
P_R \left ( {\bf l} | {\bf m} \right ) 
P_E \left ( {\bf m} |  {\bf n}^t \right )
\end{equation}
where the conditional probabilities are given by Eq.\ (\ref{P_gen}) with the
generic transition matrix replaced by the specific matrices ${\bf U}$, 
${\bf R}$ and ${\bf E}$
as indicated. Here we have used the notation 
\begin{equation}
P_T \left ( {\bf k} | {\bf l} \right ) = \prod_{j=0}^N
P_T \left ( {\bf b}_j |  l_j \right )
\end{equation}
with $k_i = \sum_j b_{ij} ~\forall i$. Moreover, using 
\begin{equation}
\sum_{{\bf k}} k_i P_T \left ( {\bf k} | {\bf l} \right ) = 
\sum_{j} T_{ij} l_j ,
\end{equation}
we find
\begin{equation}
\langle n^{t+1}_i \rangle =  \sum_{jkl} U_{ij}R_{jk} E_{kl} n_l^t ,
\end{equation}
which, by making explicit the dependence of ${\bf E}$ on ${\bf n}^t$, 
can be rewritten as
\begin{equation}\label{n_rec}
\langle n^{t+1}_i \rangle  =  \sum_{jk} U_{ij}R_{jk} 
\left \{ n_k^t \alpha_k + \frac{n_k^t}{M-1} \left [
\sum_l \left ( 1 - \alpha_l \right ) n_l^t - \left ( 1 - \alpha_k \right )
\right ] \right \} .
\end{equation}
At this stage we can readily derive a mean-field recursion equation
for the average number of demes of type $i$. In fact, assuming that
the covariance 
\begin{equation}\label{cov}
\mbox{Cov} \left ( n_i^t, n_j^t \right ) =  
\langle n_i^t n_j^t \rangle - \langle n_i^t \rangle \langle n_j^t \rangle 
\end{equation}
vanishes at any $t$ for all pairs $(i,j)$, and setting 
$\langle n_i^t \rangle =  \nu_i^t$ yield
\begin{equation}
\nu_i^{t +1}= \sum_{jk} U_{ij}R_{jk} 
\left \{ \nu_k^t  \alpha_k + \frac{\nu_k^t}{M-1} \left [
\sum_l \left ( 1 - \alpha_l \right ) \nu_l^t - \left ( 1 - \alpha_k \right )
\right ] \right \} .
\end{equation}
Thus, rather than studying the evolution 
of a specific population,  in this approximation scheme we focus on 
the evolution of an {\it average} population
whose deme frequencies at each generation are regarded as the average
of the deme frequencies of an infinite number of populations at
that generation.
Of course,  for finite $M$ the covariance can never vanish for all 
pairs $(i,j)$ since
the random variables $n_i^t$ are not statistically independent
(for instance, they obey the normalization condition $\sum_i n_i^t = M$).
However, depending on the values of the control parameters 
$\tau$, $u$ and $M$, either the covariance $\mbox{Cov}(n_k,n_l)$ or 
its coefficient
$\sum_j U_{ij}R_{jk} (1 - \alpha_l)$ may be sufficiently small  
so as to validate the mean-field equation as a good approximation.

The (conditional) covariance after a life cycle given ${\bf n}^t$
in generation $t$ is  simply given by
\begin{equation}\label{covar}
\mbox{Cov} \left ( n_i^t, n_j^t \right ) =  - \sum_{k}
S_{ik} S_{jk} n^t_{k} ~~~~~ i \neq j
\end{equation}
where  $ S_{ij} = ( U R E )_ {ij}$ is 
the matrix element of the product of the three 
transition matrices.
The case $i=j$ corresponds to
the variance, $\mbox{Var} \left ( n_i^t \right )
= \mbox{Cov} \left ( n_i^t, n_i^t \right ) $, and yields
\begin{equation}\label{varian}
\mbox{Var} \left ( n_i^t \right ) =
\sum_{k} S_{ik} \left [ 1 - S_{ik} \right ] n^t_{k} .
\end{equation}
We note that these quantities can be readily evaluated within 
the mean-field approach by replacing $n^t_{k}$ by its average,
$\nu_k^t$. Of course, the estimate of
the magnitude of the fluctuations of the random variables
$n_i~(i=0,\ldots,N)$ around their means is crucial to assess
the relevance of the finite $M$ effects.


\section{Analysis of the results}


The quantity of interest is the fraction of altruistic individuals
in the metapopulation in the stationary regime, defined as
\begin{equation}\label{p_1}
p = \frac{1}{N} ~ \sum_{i=0}^N  i \,  Y_i
\end{equation}
where $Y_i = n_i/M$ is the frequency of demes of type $i$. 
Clearly, $\sum_i Y_i = 1$. Interestingly, in
the case $i_c = N$, there is a simple relation between $p$  
and the mean fitness of the metapopulation which is defined as
$\bar{\alpha} = \sum_i \alpha_i Y_i$, namely, 
\begin{equation}\label{mean}
\bar{\alpha} = \half \left ( 1 + p \right ) .
\end{equation}
To measure the dispersion of the random variable
$p$ we introduce the variance 
\begin{equation}\label{sigma_1}
\sigma_p^2 = \frac{1}{N^2} ~ \sum_{i=0}^N i^2 \, Y_i
- p^2  
\end{equation}
which vanishes in the case of an homogeneous metapopulation
($Y_k = 1$ and $Y_i = 0$ for $ i \neq k$), and reaches
the maximum value $1/4$ in the case that the demes are 
segregated in equal proportions into the two opposite classes 
($Y_0 = Y_N = 1/2$).
Since $p$ and $\sigma_p^2$ are random variables, we will focus on their
average values, denoted by $\langle p \rangle $
and $\langle \sigma_p^2 \rangle $, respectively.
In all simulations discussed in this work, the symbols
represent the averages over $2~10^3$ independent experiments.
The error bars are calculated by measuring the standard deviation
of the average results obtained in $50$ sets of experiments, each one 
involving $40$ independent runs. Moreover,
in each run the population is left to evolve for $2~ 10^3$ generations
and  we average over the quantities under analysis  in the  
last $100$ generations. No significant
differences were found for longer runs.
Throughout our analysis
we set  $i_c = N = 10$, so that Eq.\ (\ref{mean}) holds true.

In Figs.\ 1(a) and 1(b) we present $\langle p \rangle $
and $\langle \sigma_p^2 \rangle $, respectively, as  functions
of the mutation rate $u$ for $\tau = 0.9$ and two representative
values for the number of demes,  $M=10$ and $M=100$.
For $u=0$ the altruistic trait always takes
over the metapopulation, provided that there is at least one altruistic
deme in the initial state ($n_N^0 \geq 1$). 
Besides the stable fixed point presented in these figures,
for large $\tau$ the mean-field recursion equations possess an unstable 
one, $p = u$, which can be reached by starting the iteration
with  non-altruistic demes only ($n_0^0 = M$).
Thus the effect of finite $M$, as shown  by the results of the
simulations, is to increase the instability of the altruistic regime 
against mutations by stabilizing the mean-field unstable fixed point. 
(We note that the mean-field results actually shown the opposite tendency, 
which indicates the failure of the approximation in this matter.)
This is expected since in a smaller metapopulation, chance plays a 
greater role, and so deleterious mutations accumulate with a higher probability,
causing a more rapid decrease in the mean fitness of the metapopulation.
In the case that the size of the metapopulation  is not fixed but
depends on its mean fitness, this  positive feedback, 
termed mutational meltdown,  leads rapidly to the extinction of
the metapopulation \cite{Gabriel,Bezzi}. 
The agreement between the mean-field predictions and
the simulations are very good for $M=100$, except in the region just after the 
variance maximum. Up to this maximum the population is
composed almost exclusively of altruistic and non-altruistic demes, while
beyond it the
number of altruistic demes decreases very rapidly, the sole source
of altruistic individuals being the mutations within 
the non-altruistic demes. Clearly, in this scenario
we have  $p = u$ in agreement with the simulation results. 
The occurrence of a pronounced maximum in $\sigma_p^2$
indicates the existence of a phenomenon similar to the error
threshold transition of Eigen's quasispecies model for molecular evolution 
\cite{Eigen}. (The formal similarity between group selection and 
molecular evolution models has already been pointed out in ref.\ 
\cite{Ana}.) We note that even for $M=10$, the mean-field approximation 
yields very good  results for small $u$. Of course, the agreement between 
theory and simulations 
is more problematic for small $M$, since in this case the probability that the 
altruistic demes are  lost from the metapopulation due 
solely to fluctuations becomes
significant, leading to the so-called stochastic escape phenomenon \cite{Higgs,Alves}.
This loss is practically irreversible as the altruistic selective 
disadvantage is too high to allow for the production of new altruistic demes.

The same quantities,   $\langle p \rangle $
and $\langle \sigma_p^2 \rangle $, are presented in Figs.\ 2(a) and 2(b), 
except that the altruistic disadvantage is reduced to $\tau = 0.2$.  In this
case, the effects of the finite $M$ fluctuations are almost suppressed
as illustrated by the agreement between the mean-field and the simulation
results. The stochastic escape phenomenon is not important in this
case since new altruistic demes can readily be generated due to the 
small reproductive disadvantage of the altruistic individuals. 
The results
for $M\rightarrow \infty$ are practically indistinguishable from
those obtained in the mean-field approximation for $M =100$. We
have verified that changing the values of $i_c$ and $N$ alters the
frequency of the altruistic gene smoothly, leaving its qualitative
dependence on the mutation rate unaffected.

A more direct measure of the finite $M$ fluctuations is
presented in Figs.\ 3 and 4, which show
the variances of the fraction  of non-altruistic and altruistic
demes, $\mbox{Var} \left ( Y_0 \right )$ and 
$\mbox{Var} \left ( Y_N \right )$, 
respectively, as functions of the mutation rate for $\tau = 0.9$. 
We note that although the mean-field approximation describes very 
well the fluctuations outside the region where 
the transition between the altruistic ($p \approx 1$) and non-altruistic
($p \approx u$) regimes 
takes place,
it fails badly in that region. This failure seems  more pronounced
in Fig.\ 4 because the range of $u$ coincides with the transition region, 
but a similar discrepancy occurs in Fig.\ 3 also, where the variance peaks 
for small $u$ are
completely overlooked by the mean-field approximation. 
The situation for $\tau = 0.2$ is illustrated in  Fig.\ 5 where
we present $\mbox{Var} \left ( Y_N \right )$ as a function of $u$.
In this case the mean-field approximation reproduces  very well 
the behavior pattern of the simulation results, except for the 
heights of the peaks which, as expected, are underestimated.
The results for $\mbox{Var} \left ( Y_0 \right )$ are  very
similar to those shown in Fig.\ 5. 
Of course, the disagreement between simulation and theory
is expected since we are trying to estimate the size
of the fluctuations using an approximation scheme that  neglects those very
same fluctuations. However, the surprisingly good agreement shown in Fig.\ 3
for $u$ outside the transition region suggests that a self-consistent 
iterative scheme, where the covariances calculated in the mean-field approximation
are used to improve that approximation, may describe successfully the finite 
$M$ fluctuations. 
As expected, these variances tend to zero as the number of demes 
increases.


\section{Conclusion}


In this paper we have modified the standard implementation 
of the group selection mechanism, which considers first the extinction 
of the demes and then the recolonization of the extinct  demes by the 
surviving ones \cite{Levins,Kilmer,Aoki,Ana}, by assigning  a
recolonizing deme to each extinct deme  immediately 
after its extinction, according to Eq.\ (\ref{sur}). 
The actual replacement of the extinct demes is carried out 
simultaneously for all demes following the stochastic prescription
given in Eq.\ (\ref{P_gen}). This modified extinction-recolonization 
procedure avoids the otherwise inevitable global extinction of
the population. We have verified, however, that this procedure
yields qualitatively similar results to those 
obtained with the standard extinction-recolonization procedure in 
the case that the metapopulation survives global extinction long
enough to reach a metastable state.

It is important to  mention that, in contrast 
to its original and very criticized ecological motivation 
\cite{Wynne,Wilson}, some concepts borrowed from  group selection 
have  been successfully applied to describe the evolution of
parasite-host systems \cite{Pimentel,Frank}. In this
case the hosts are associated with  the demes, while the parasites
correspond to the individuals inhabiting the demes. 
The role of the altruistic individuals is played by the less virulent
parasites which, by having a lower growth rate, increase the survival 
probability of the host. Migration of individuals
between demes corresponds to horizontal transmission of parasites.
The transmission of the parasite between parent and offspring 
generations is termed vertical transmission. Interestingly,
a well-known result is that, in
a population of  asexual hosts, parasites with vertical transmission 
alone cannot persist if the infected hosts suffer any fitness 
cost \cite{Fine,Lipsitch}. (This result is readily recognized as 
Eshel's \cite{Eshel}, although no reference to that author is made
in the specialized literature of parasite-host systems.) 
Our finding that at certain ranges of the
mutation rate (around $0.04$ in Fig.\ 1), virulent parasites with
vertical transmission alone almost take over the
population yields evidences of the major role played by mutations
in the evolution of virulence \cite{Now}. This dominance becomes
more pronounced as the host population decreases.
A more thorough formulation of parasite-host dynamics through the 
classical, discrete time  population genetics formalism 
used to study group selection models is still 
lacking. Such formulation will certainly help to uncover many more 
similarities, as well as overlapping results, between these two
fascinating research fields.


\begin{ack}
This work was supported in part by Conselho Nacional de Desenvolvimento
Cient\'{\i}fico e Tecnol\'ogico (CNPq).
\end{ack}



\newpage

\section*{Figure captions}

{\bf Fig.\ 1(a)} Average frequency of altruists as  function
of the mutation rate for $\tau = 0.9$, $M = 10 \left( \triangle
\right ) $,  and $M = 100 \left( \bigcirc \right ) $. The solid 
and dashed curves are the mean-field results for $M=10$ and
$M=100$, respectively. The straight line is $\langle p \rangle =u$.  
The error bar is omitted when
it is smaller than the symbol size. The parameters are
$i_c = N = 10$.

{\bf Fig.\ 1(b)} Average variance of the frequency of altruists 
as  function of the mutation rate. The parameters and convention
are the same as for Fig.\ 1(a).

{\bf Fig.\ 2(a)} Same as Fig.\ 1(a) but for $\tau = 0.2$.

{\bf Fig.\ 2(b)} Same as Fig.\ 1(b) but for $\tau = 0.2$.

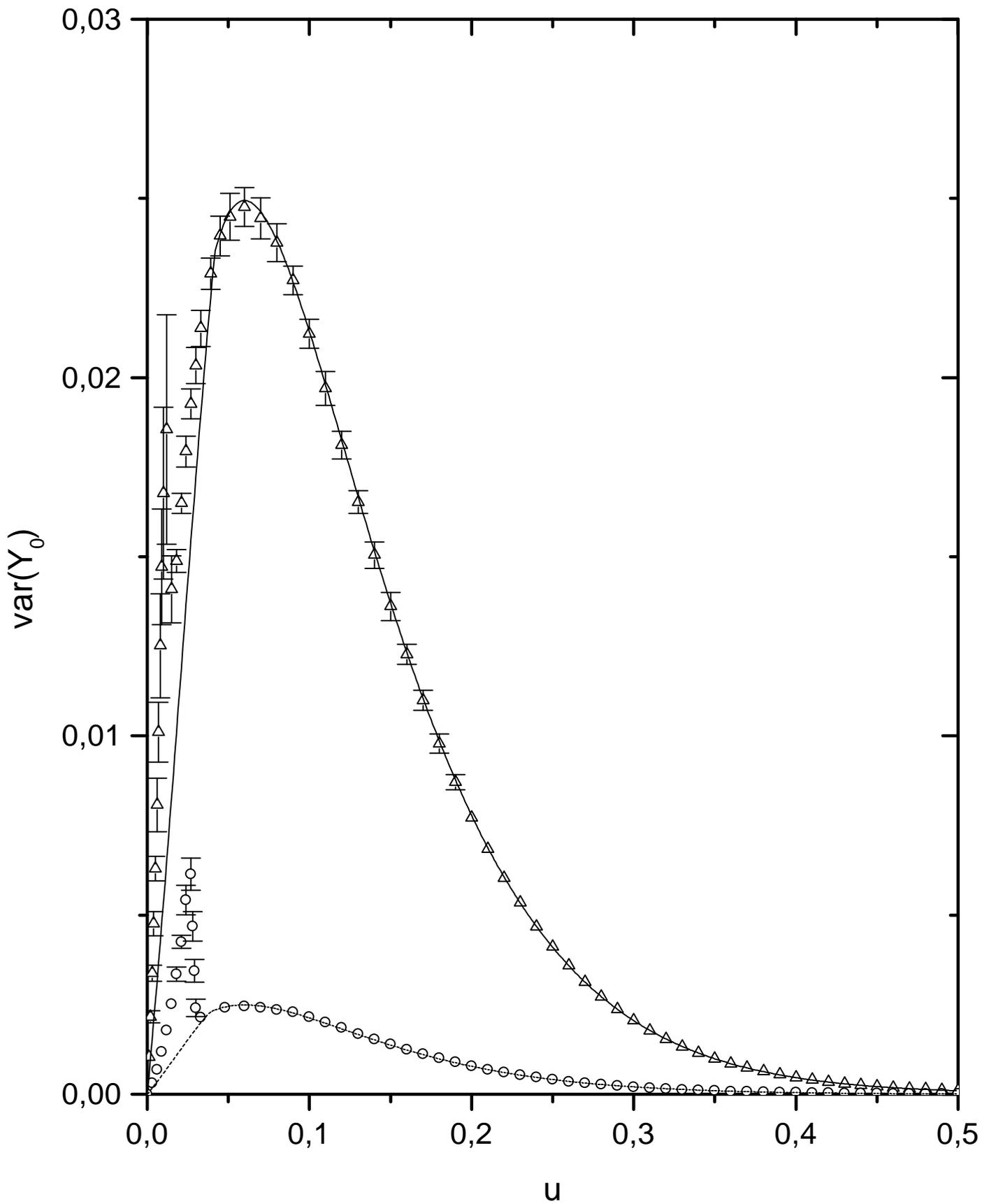
{\bf Fig.\ 3} Variance of the fraction  of non-altruistic 
demes, $\mbox{Var} \left ( Y_0 \right )$, 
as function of the mutation rate for $\tau = 0.9$. The convention
is the same as for  Fig.\ 1(a).

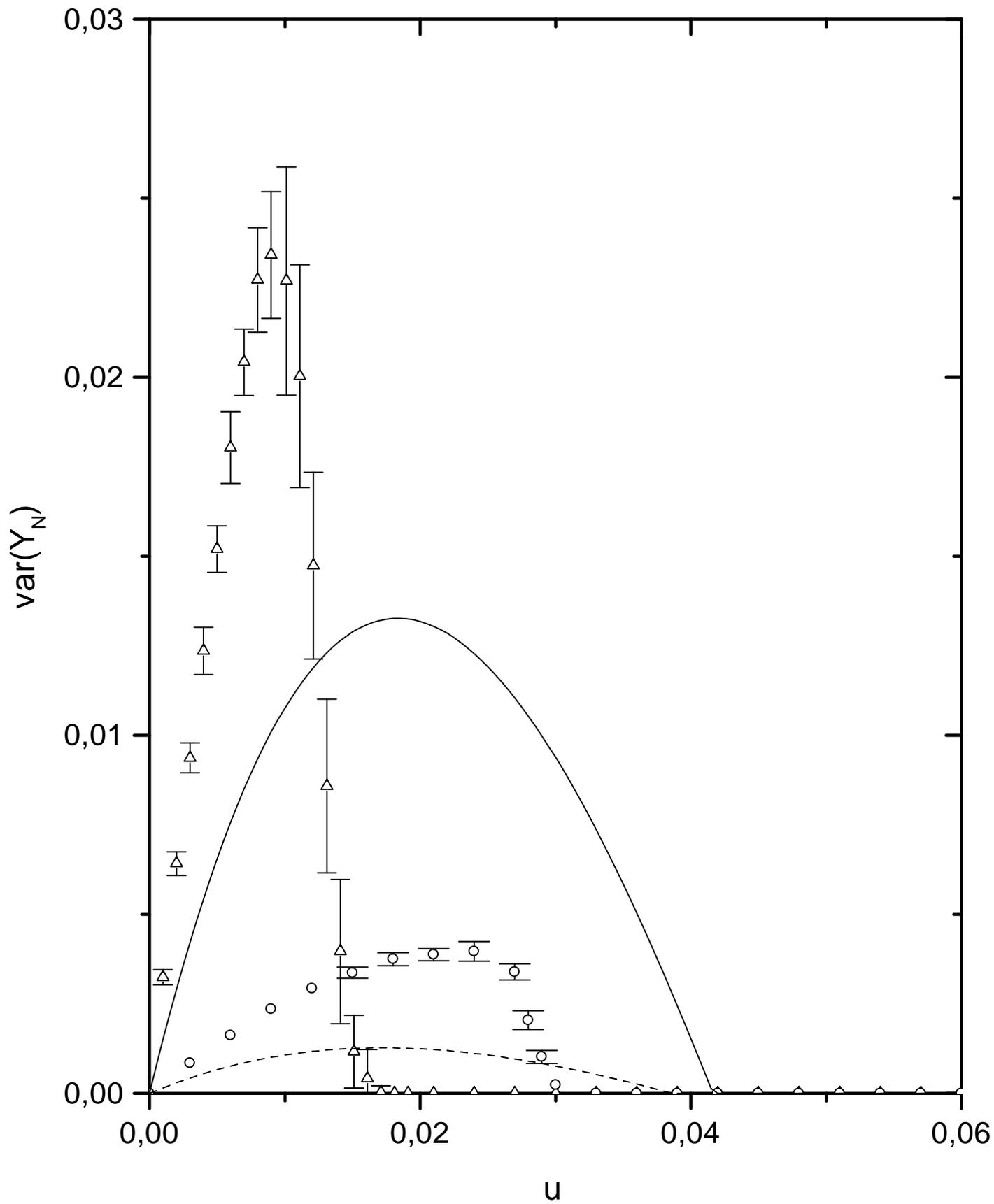
{\bf Fig.\ 4} Variance of the fraction of altruistic demes,
$\mbox{Var} \left ( Y_N \right )$,
as function of the mutation rate for $\tau = 0.9$.
The convention is the same as for  Fig.\ 1(a).

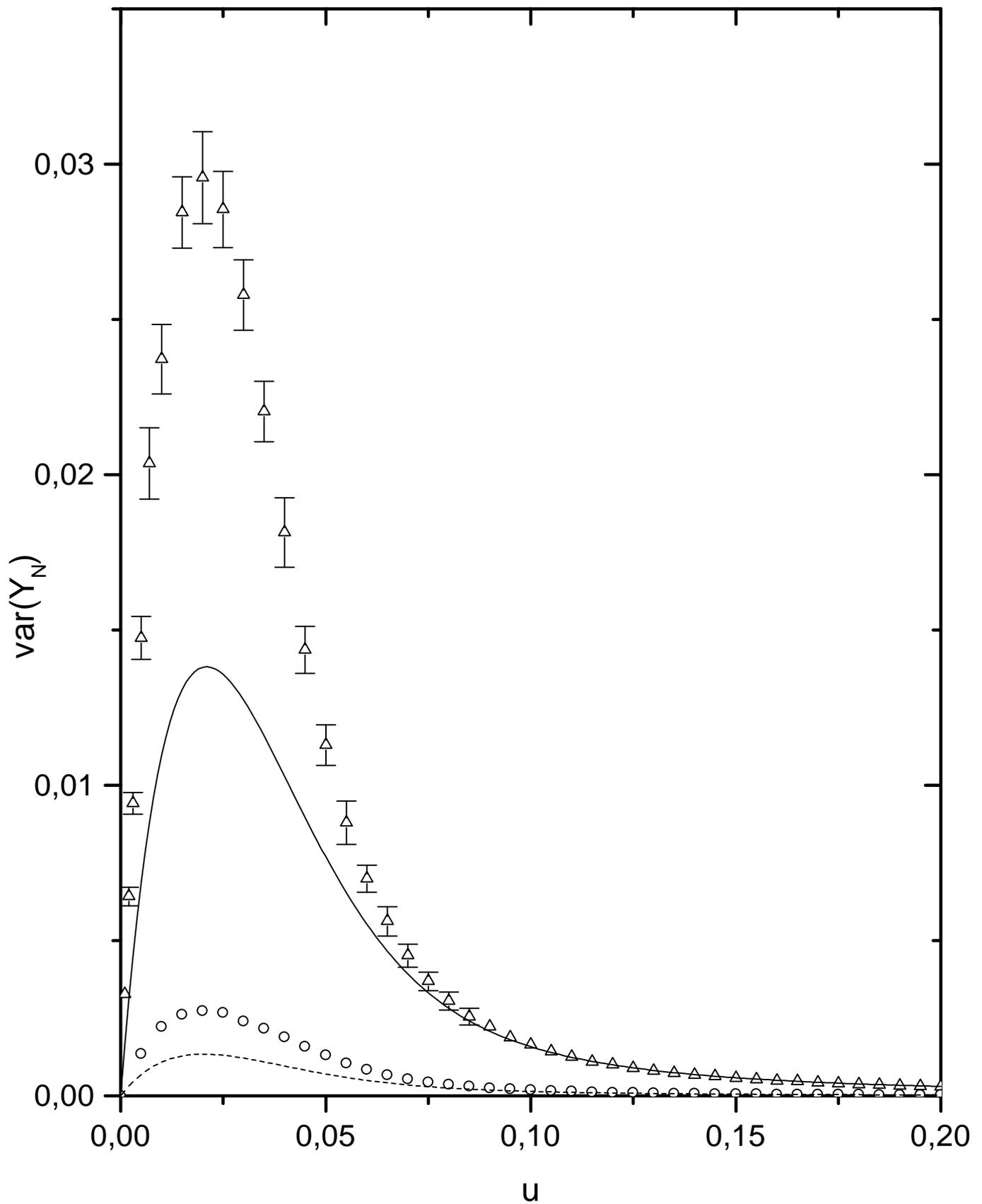
{\bf Fig.\ 5} Same as Fig.\ 4, but for $\tau = 0.2$.

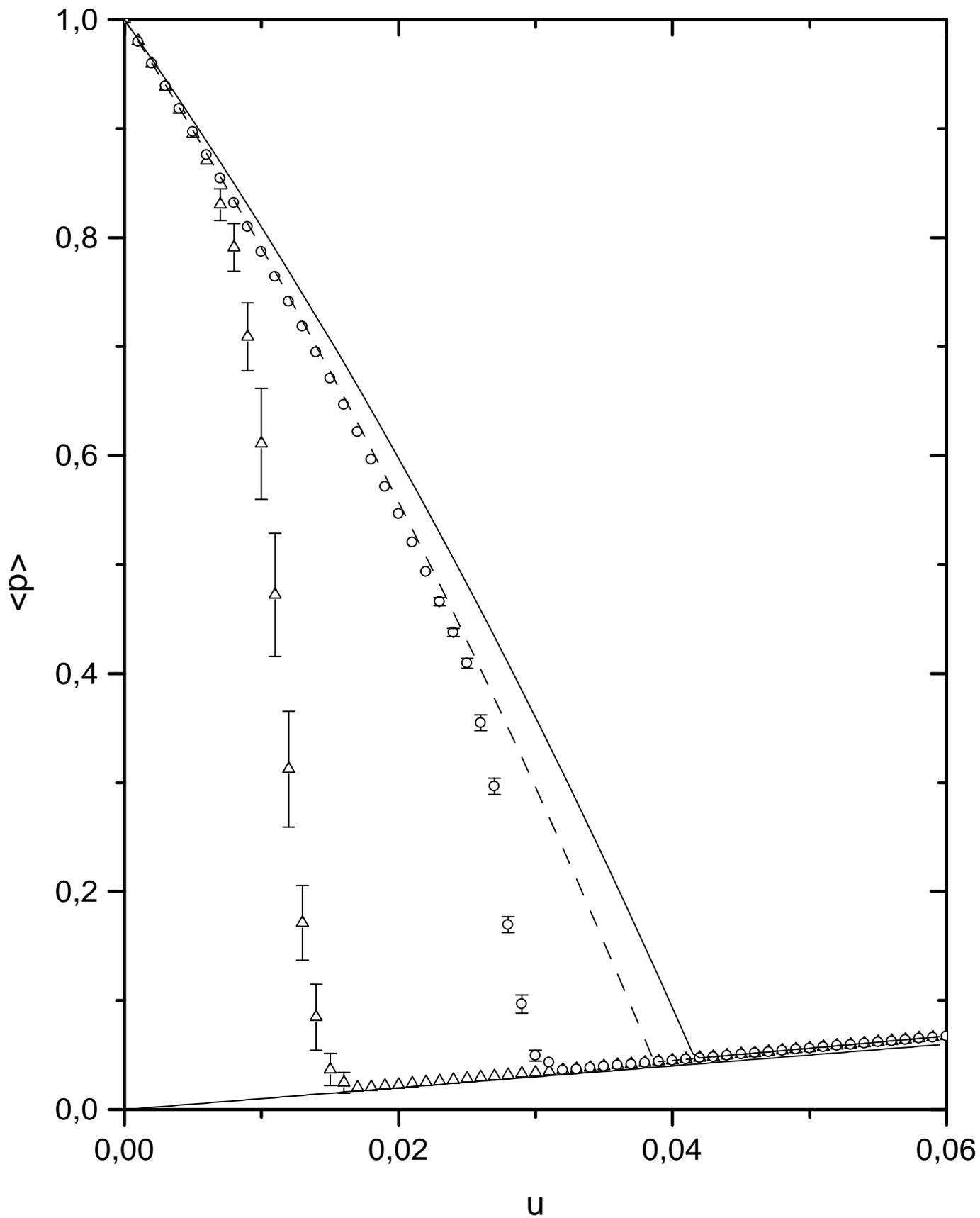
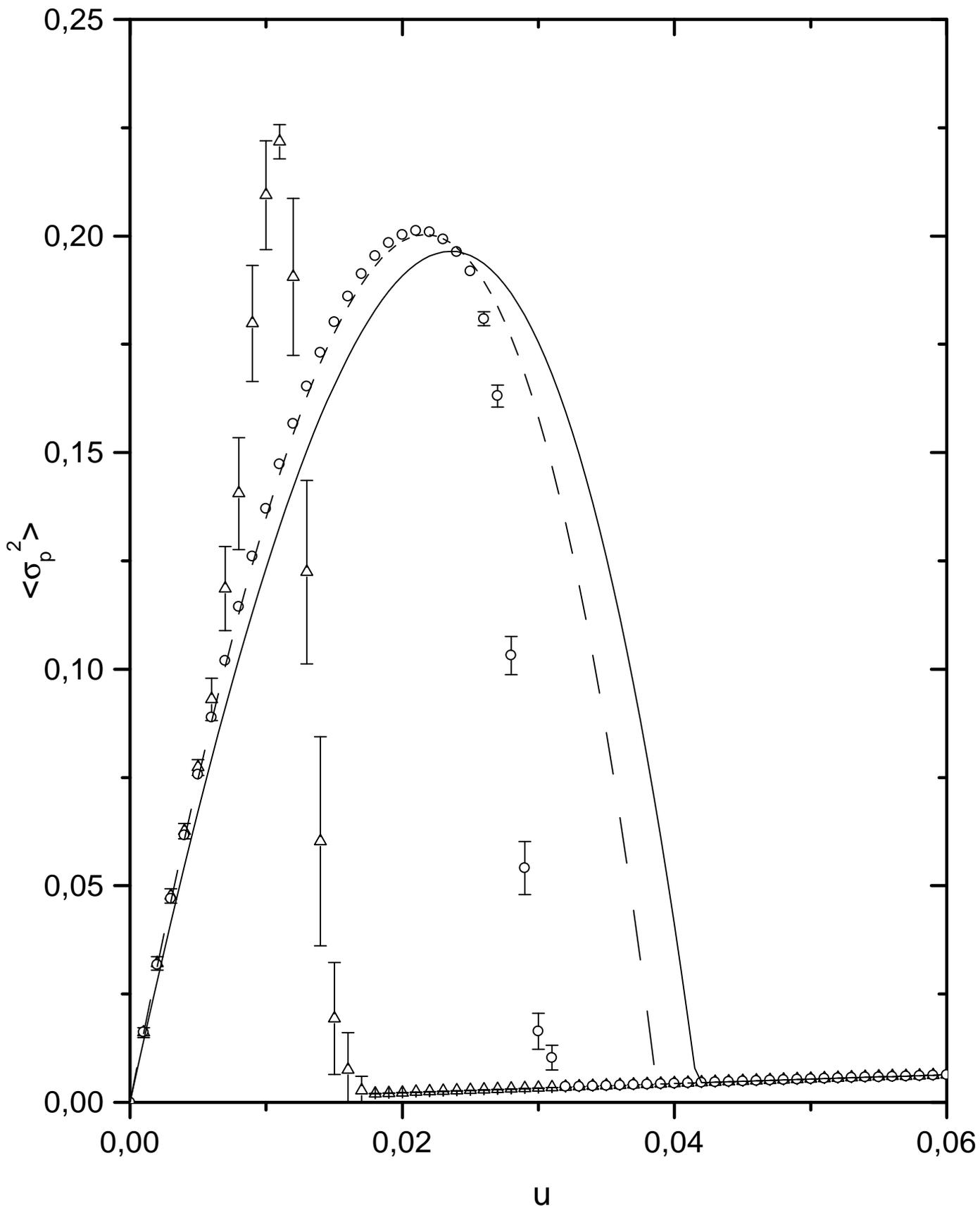
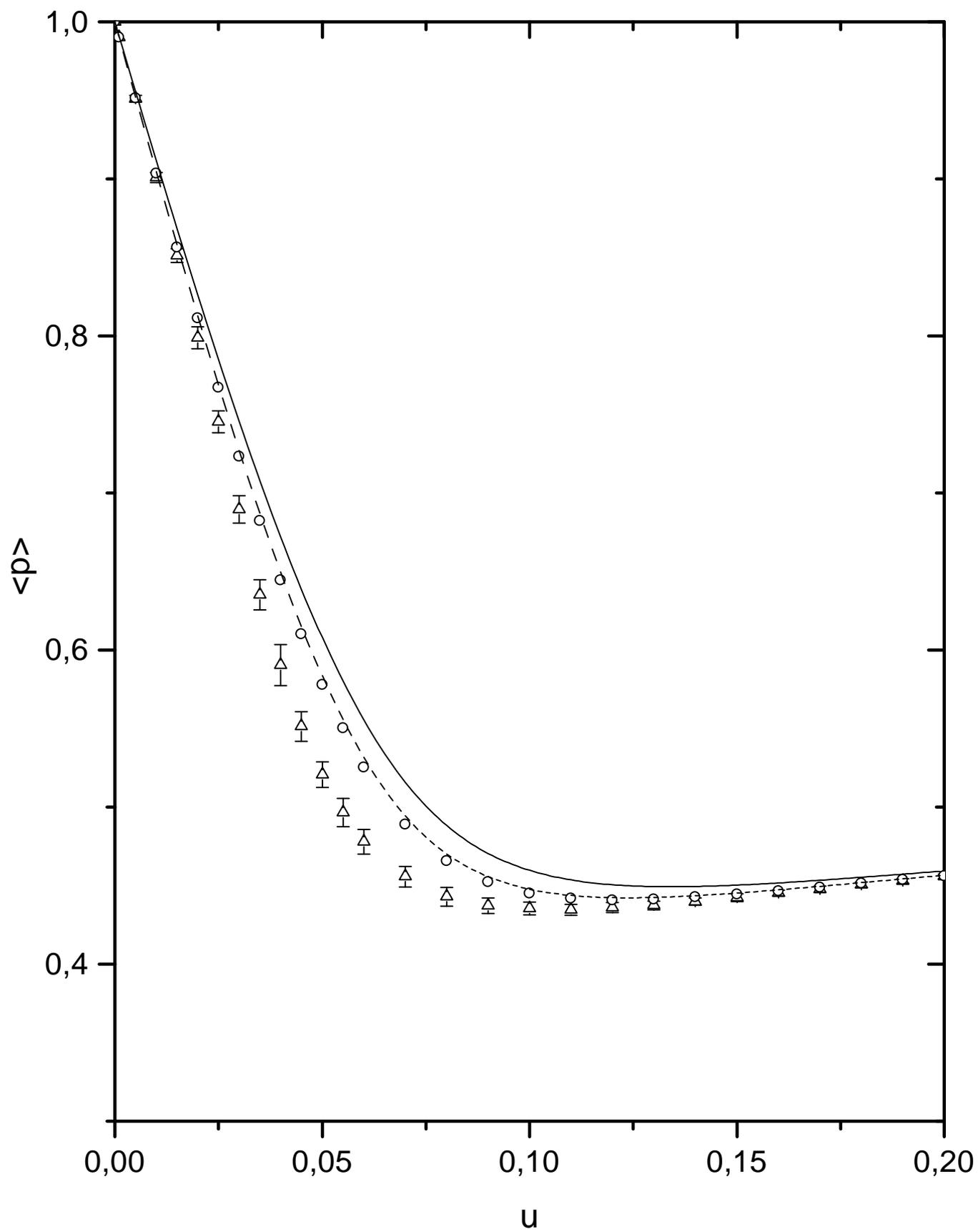
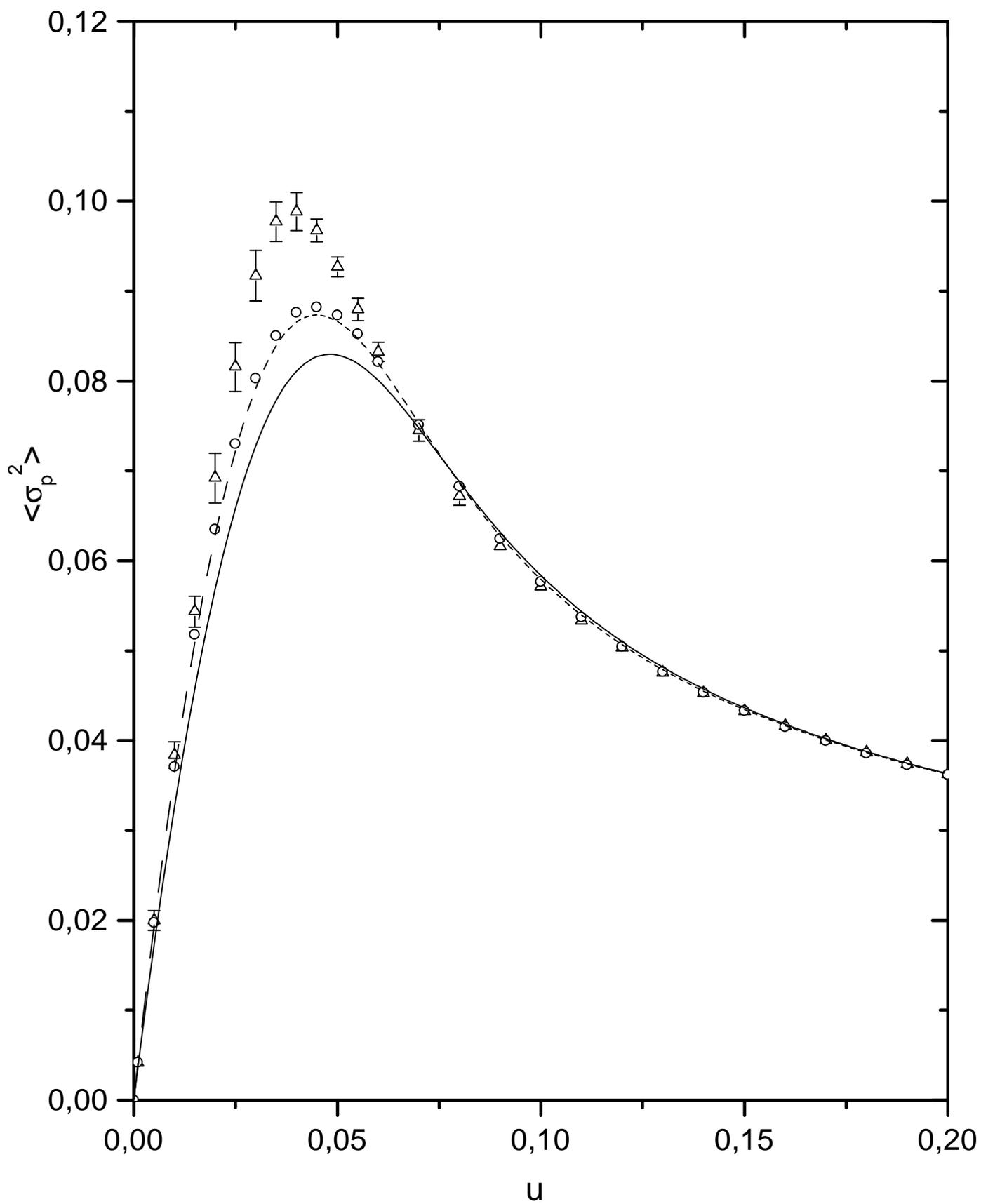
\end{document}